\documentclass[pdflatex,sn-mathphys]{sn-jnl}

\usepackage{amsmath}
\usepackage{subfig}
\usepackage{graphicx}
\usepackage{booktabs}

\jyear{2023}%

\theoremstyle{thmstyleone}%
\theoremstyle{thmstyletwo}%

\theoremstyle{thmstylethree}%

\unnumbered

\begin{document}

\title[ ]{A switched auxiliary loss for robust training of transformer models for histopathological image segmentation}


\author*[1]{\fnm{Saharsh} \sur{Barve}}\email{saharsh@onwardhealth.co}
\equalcont{These authors contributed equally to this work.}

\author[2]{\fnm{Mustaffa} \sur{Hussain}}\email{mustaffa@onwardhealth.co}
\equalcont{These authors contributed equally to this work.}

\author[3]{\fnm{Mohnish} \sur{Pakanati}}\email{mohnish@onwardhealth.co}

\affil*[1]{\orgname{Onward Assist}, \orgaddress{\street{}, \city{Bangalore}, \postcode{560066}, \state{Karnataka}, \country{India}}}


\abstract{Functional tissue Units (FTUs) are cell population neighborhoods local to a particular organ performing its main function. The FTUs provide crucial information to the pathologist in understanding the disease affecting a particular organ by providing information at the cellular level. In our research, we have developed a model to segment multi-organ FTUs across 5 organs namely: the kidney, large intestine, lung, prostate and spleen by utilizing the 'HuBMAP + HPA - Hacking the Human Body' competition dataset. We propose adding shifted auxiliary loss for training models like the transformers to overcome the diminishing gradient problem which poses a challenge towards optimal training of deep models. Overall, our model achieved a dice score of 0.793 on the public dataset and 0.778 on the private dataset. The results supports the robustness of the proposed training methodology. The findings also bolster the use of transformers models for dense prediction tasks in the field of medical image analysis. The study assists in understanding the relationships between cell and tissue organization thereby providing a useful medium to look at the impact of cellular functions on human health.
}




\maketitle

\section{1. Introduction}\label{sec1}

The creation of a Human Reference Atlas (HRA), a detailed map of the healthy adult human body at the cellular level, is a crucial endeavor in medical research. A key component of this atlas is the segmentation of Functional Tissue Units (FTUs). These FTUs represent the smallest building blocks within an organ that perform specific physiological functions. They bridge the gap between whole organs and individual cells, providing a vital level of detail for studying human biology. Accurate segmentation of FTUs from medical images, often through machine learning techniques, is essential for characterizing their properties, distribution, and variations across the human population. This information will ultimately contribute to a comprehensive HRA with the potential to revolutionize our understanding of health and disease. 

That said, a major hurdle in this quest is the time-consuming and expensive nature of manual FTU segmentation. It simply can't handle the ever-growing volume of medical data. Consider the lungs, for example. These complex organs are estimated to contain millions of alveoli (a type of FTU) responsible for gas exchange. Even a trained pathologist would require a significant amount of time to manually segment just a small portion of lung tissue. This illustrates the difficulties in analyzing large-scale FTUs of organs using manual methods.

The HuBMAP + HPA Hacking the Human Body competition brought together a global community of data scientists, researchers, and enthusiasts. A total of 1175 teams from 78 countries participated in this challenging event. The primary objective was to develop robust models for segmenting functional tissue units (FTUs) in tissue section images from five different organs: kidney, large intestine, spleen, lung, and prostate. The training dataset was sourced from the Human Protein Atlas (HPA), while the hidden test dataset came from the Human BioMolecular Atlas Program (HuBMAP). The test data had different image parameters, including pixel size, tissue thickness, and staining protocol, necessitating models that could generalize effectively. The evaluation metric was the dice coefficient, calculated on the hidden test set.

Convolutional neural networks(CNNs) have been at the forefront of computer vision for tasks like image classification \cite{cnn_classify_1,cnn_classify_2}, image segmentation \cite{cnn_segment_1,cnn_segment_2}, and object detection \cite{cnn_obg_detect_2,cnn_obj_detect_1}. The applications of such deep learning-based models have proved to be crucial in the recent advancements made in the medical field toward intelligent system-aided predictive analytics \cite{cnn_med_2,cnn_med_1}. 
Recently, several researchers have suggested that the transformer encoder models have matched and even outperformed CNNs models in several of the state-of-the-art vision tasks. With the success transformer architectures have achieved, their incorporation into clinical medical practices has tremendous potential in the near future. One key advantage of transformers is their inherent ability to model long-range dependencies within data. This is particularly valuable in medical applications where subtle contextual information in images can be critical for diagnosis and treatment planning. While these models have shown promise in FTU segmentation tasks, as evidenced by competitions like 'HuBMAP + HPA - Hacking the Human Body', they can still face challenges for optimal training due to vanishing gradients and infinitesimal losses during backpropagation. This phenomenon hinders the model's ability to learn effectively, especially in deeper architectures, and can ultimately affect its generalizability and robustness – qualities essential for reliable medical applications.

To address this challenge, we propose a novel approach that incorporates switched auxiliary loss during model training. This method aims to mitigate the vanishing gradient problem and improve the overall training efficiency of transformer models.  We implement the same on 3 transformer models, namely, CoaT \cite{coat}, Pyramid Vision Transformer-v2 \cite{pvtv2} and Segformer \cite{segformer}, in the FTU binary segmentation task. We further implement our approach on common CNN architectures used in the medical domain such as Unet \cite{unet} and DeepLabv3+\cite{deeplabv3plus} report our findings.

\section{2. Literature Review}\label{sec3}
Convolutional neural networks (CNNs) are currently the most widely used approach for image classification tasks in computer vision. However, recent developments in transformer architectures have shown promising results in this field. Vision Transformer (ViT), introduced by Dosovitskiy et al. \cite{vit}, is one such transformer architecture that uses multi-head self-attention to process image patches directly and has achieved competitive performance on image classification benchmarks. Despite its success, ViT produces a single-scale low-resolution feature map, which limits its effectiveness when it comes to dense prediction tasks like medical image segmentation. To address this, Wang et al. proposed Pyramid Vision Transformer (PVT) and PVTv2 \cite{pvt,pvtv2}, which utilize a hierarchical structure to generate multi-scale feature maps suitable for segmentation tasks. Enze Xie et al. \cite{segformer} have developed Segformer, a feed-forward network that addresses the issue of single-scale outputs in computer vision applications. Unlike ViT and PVT architectures, Segformer uses 3x3 convolutions instead of positional encoding for better results.

Another recent innovation is the Swin Transformer by Ze Liu et al. \cite{swintransformer}. This architecture employs a window-based self-attention mechanism that leads to significant efficiency gains. Weijian Xu et al. \cite{coat} propose CoaT, which incorporates co-scale and conv-attentional mechanisms, achieving high accuracy and efficiency on image recognition tasks. The lightweight characteristic of transformer architecture makes it suitable for working with large medical image datasets. This has led to its application in solving tasks like Brain Tissue Image segmentation  \cite{Jinjing-brain_tissue}, and myocardial fibrosis tissue detection \cite{Yuhan-fibrosis} to name a few. Additionally, new models have also been proposed specifically for medical image analysis, by combining the transformer's encoding capabilities with Unet architecture, such as TransUNet \cite{trans_unet}, Mix Transformer-Unet \cite{mix_transformer_unet}. These hybrid architectures aim to leverage the strengths of both CNNs (efficient local feature extraction) and transformers (global dependency modeling) to achieve superior performance in medical image segmentation tasks \cite{dualswin, gao2021utnet}. Ongoing research focuses on the development of transformer architectures specifically designed for medical image analysis. These architectures incorporate domain-specific knowledge into the models \cite{du2023mdvit} and aim to develop more efficient transformer variants for processing medical imaging data \cite{liu2022swin}. These advancements have the potential to improve the accuracy, efficiency, and generalizability of medical image segmentation tasks.

The auxiliary loss was initially proposed in the Inception model \cite{inception} with the motivation to overcome the problem posed by diminishing gradients. However, with the introduction of deep architectures such as transformers, Dasol Han et al. \cite{Han_dasol-aux_loss_1} proposed a rebirth of this technique for stable and regularized training of the model. Yechan Yu et al. \cite{Yechan-aux_loss_2} applied auxiliary loss to each encoder block to increase model training and improve accuracy. The effectiveness of auxiliary loss extends beyond transformers. Ji et al. \cite{Yuanfeng-aux_loss_3} leverage it in their MC-Transformer architecture to facilitate "proxy-embedded learning," where intermediate representations serve as proxies for the final task. Overall, auxiliary loss has become a widely-used technique for training complex neural networks. Recent research has explored its effectiveness in various contexts. For instance, Luo et al. \cite{Luo_aux_loss_4} have demonstrated that it can be used to reduce overfitting in deep metric learning tasks. Similarly, Jiao et al. \cite{Jiao_aux_loss_5} have used auxiliary losses to improve the convergence and generalization of deep reinforcement learning models.

\subsection{3. Exploratory Dataset Analysis (EDA)}\label{subsec4.1}

\begin{figure}%
\centering
\subfloat[\centering  Prostate]{{\includegraphics[width=0.2\textwidth]{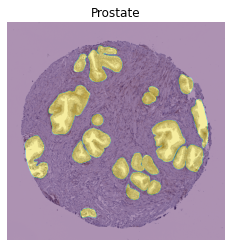} }}%
\subfloat[\centering Spleen]{{\includegraphics[width=0.2\textwidth]{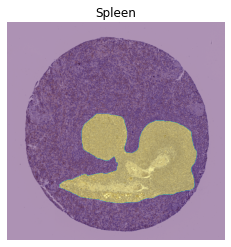} }}%
\subfloat[\centering Lung]{{\includegraphics[width=0.2\textwidth]{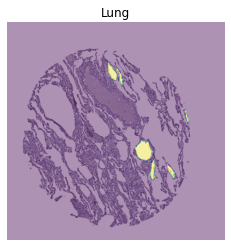} }}%
\subfloat[\centering Kidney]{{\includegraphics[width=0.2\textwidth]{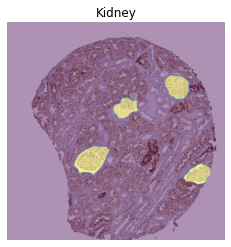} }}%
\subfloat[\centering Large Intestine]{{\includegraphics[width=0.2\textwidth]{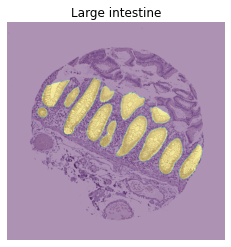} }}%
\\
\smallskip
\caption{Representation of Organ-Specific Overlays from the Training Dataset}
\label{fig:train_data}
\end{figure}

The competition provides a challenging dataset of biopsy slide images from five organs prostate, spleen, lung, kidney and large intestine (Figure \ref{fig:train_data}). The goal is to identify and segment functional tissue units (FTUs) within these images. The data originates from two sources: Human Protein Atlas (HPA) and Human BioMolecular Atlas Program (HuBMAP).  Images come from diverse sources and are prepared using different protocols at varying resolutions, reflecting real-world complexities in medical image analysis. The training set uses public HPA data, while the public and private test sets incorporate a mix of private HPA and HuBMAP data.  
A key challenge lies in developing models that can adapt and perform well on unseen data prepared with different protocols, promoting generalizability and real-world applicability. Table \ref{table:pixel_sizes} highlights the differences in pixel sizes between HPA and HubMAP. Across all organs, HPA consistently reports a pixel size of 0.4 micrometers, whereas HubMAP's data shows a wider range, from 0.229 to 6.263 micrometers. While the HPA utilized a consistent slice thickness of 4 µm for all images, the HuBMAP dataset presents a varied range, with tissue slice thicknesses ranging from 4 µm (for the spleen) to 10 µm (for the kidney). Notably, 351 images are in the training data, accompanied by additional properties of the respective image CSV format such as image label (HPA or HuBMAP), organ, height and width dimensions, pixel sizes, tissue thicknesses, age, and sex. The HPA samples underwent staining with antibodies, which were then visualized using 3,3'-diaminobenzidine (DAB). They were then counterstained with hematoxylin. In contrast, images from HuBMAP were prepared using Periodic acid-Schiff (PAS)/hematoxylin and eosin (H\&E) stains. Additionally, there is an imbalance in the number of samples across the organs, depicted in Figure \ref{fig:train_data}.  The total size of the data is 9.39GB.
\begin{table}[h]
    \centering
    \caption{Comparison of Pixel Sizes in Different Organ Datasets}
    \begin{minipage}{0.7\textwidth} 
    \centering
    \begin{tabular}{lcc}
        \toprule
        \textbf{Organ} & \textbf{HPA Pixel Size} & \textbf{HubMAP Pixel Size} \\
        \midrule
        Kidney          & 0.4                     & 0.229                      \\
        Large Intestine & 0.4                     & 0.7562                     \\
        Lung            & 0.4                     & 0.4945                     \\
        Prostate        & 0.4                     & 6.263                      \\
        Spleen          & 0.4                     & 0.4945                     \\
        \bottomrule
    \end{tabular}
    \\
    \smallskip
    \small Note: Pixel sizes are expressed in micrometers.
    \end{minipage}
    \label{table:pixel_sizes}
\end{table}

\begin{figure}[h]
    \centering
    \includegraphics[width=0.8\textwidth]{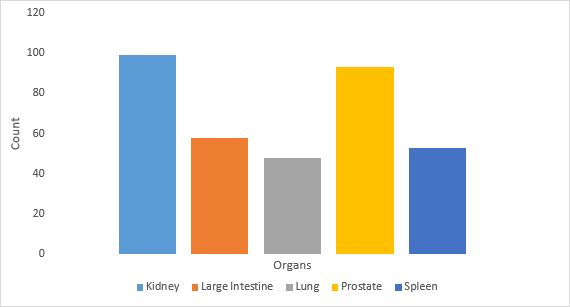} 
    \caption{The distribution of organ classes within the training dataset}
    \label{fig:example_figure}
\end{figure}
\section{4. Proposed Methodology}\label{sec4}
In this study, we have experimented with various deep learning architectures to tackle the segmentation task, where each model offers distinct features and characteristics. In addition, we added a switched auxiliary loss during the training phase to add robustness to the model.  This through method allowed us to determine how each loss formulation affected the model's ability to perform segmentation.

\subsection{4.1 Models}\label{sec4.1}

\subsubsection{4.1.1 CoAt (Convolution Attention)}\label{sec4.1.1}
CoAt, which stands for Convolutional Attention \cite{coat}, is a powerful combination of convolutional and self-attention mechanisms designed to capture both spatial and contextual information in medical images. What sets CoAt apart is its ability to handle diverse input resolutions and adapt to varying scale factors, making it an ideal choice for multi-scale feature extraction in medical imaging tasks.

\subsubsection{4.1.2 Pyramid Vision Transformer-v2 (PVTv2)}\label{sec4.1.2}
The Pyramid Vision Transformer-v2 (PVTv2) \cite{pvtv2} is a type of artificial intelligence architecture that features a hierarchical, pyramid-like design. This design enables the effective extraction of features across multiple scales. PVTv2 uses a combination of patch-based processing and token-wise attention mechanisms, which allows it to capture both local and global context information. As a result, it is an excellent choice for medical image segmentation tasks that require subtle spatial understanding.

\subsubsection{4.1.3 Segformer}\label{sec4.1.3}
Segformer \cite{segformer} is a new method for semantic segmentation that uses transformer-based architectures within a fully convolutional framework. It treats the segmentation task as a per-pixel classification problem, utilizing self-attention mechanisms to capture long-range dependencies and contextual information. This allows for more precise tracing of anatomical structures in medical images.
\smallskip
\\
In addition to these Transformer models, we incorporated CNN architectures to provide a comparative benchmark. We have used EfficientNet-B5 coupled with Deeplabv3++ and ResNet101 paired with Deeplabv3++. These CNN models served as reference points, facilitating a comprehensive evaluation of segmentation performance across different models.

\subsubsection{4.2 Metric}\label{subsubsec4.2}
We use the mean Dice coefficient metric for evaluation. The Dice coefficient can be used to compare the pixel-wise agreement between a predicted segmentation and its corresponding ground truth. The formula is given by:
\begin{equation}
\frac{2* X \cap Y}{ X + Y }  
\end{equation}

where X is the predicted set of pixels and Y is the ground truth.

\subsubsection{4.3 Losses}\label{subsubsec4.3}

The selection of the loss function is crucial in guiding model training and optimizing performance. This subsection covers three key loss functions used in our study: binary cross-entropy \cite{binary_cross_entropy}, auxiliary loss \cite{inception}, and switched auxiliary loss \cite{transformer}. 
 BCE loss calculates the difference between predicted probabilities and actual labels for each pixel. This helps the model classify pixels into foreground or background classes more accurately. Mathematically, BCE loss is defined as,
\[
\text{BCE}(y, \hat{y}) = - \frac{1}{N} \sum_{i=1}^{N} \left(y_i \log(\hat{y}_i) + (1 - y_i) \log(1 - \hat{y}_i)\right)
\]

where \(y\) represents the ground truth labels, \(\hat{y}\) represents the predicted probabilities, and \(N\) denotes the total number of samples.
\\
BCE is used to compute the auxiliary loss between the intermediate output at the end of the second block of the encoder against the down-sampled mask to the corresponding size. This loss is added along with the main loss and is used for backpropagation, as shown below,

\[
\text{Auxiliary Loss} = \text{BCE} + 0.2 \times \text{loss2}
\]

where BCE is the main loss and "loss2" represents the loss at the end of the second block of the encoder.
\smallskip

During our training, we shift the auxiliary loss from the second block intermediate output to the first encoder block intermediate output at the end of 60 epochs and further train the model for 60 more epochs.

\[
\text{Switch Auxiliary Loss} = \text{BCE} + 0.2 \times \text{loss1}
\]
where BCE is the main loss and "loss1" represents the loss at the end of the first block of the encoder.
\section{5. Training Pipeline}\label{subsubsec5}

\begin{figure}[h]%
\centering
\includegraphics[width=0.9\textwidth]{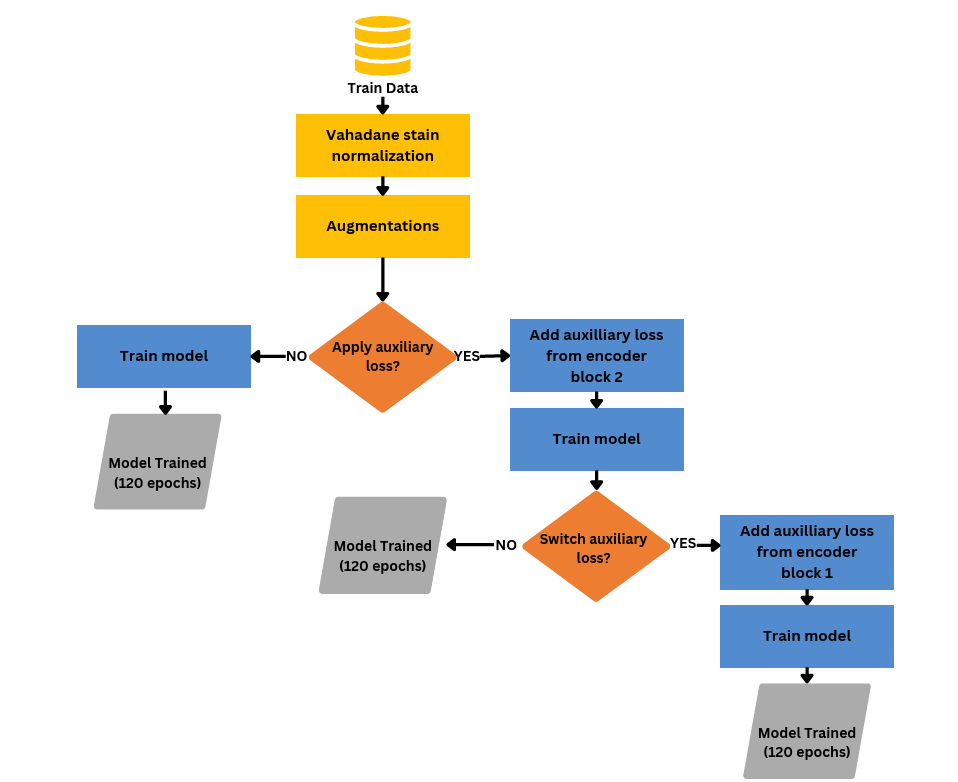}
\caption{Model training Framework}
\label{fig:fig_train}
\end{figure}

For this research, we have utilized Tesla P100-PCIE-16GB GPU with 16GB GDDR6 VRAM for model training and inference. The CPU used is a 1xcore hyper-threaded Xeon Processors @2.2Ghz with 15.26GB RAM and disk storage of 155GB.

Our training pipeline, Figure \ref{fig:fig_train}, contains several steps to ensure effective FTU segmentation.
To begin with, we performed stratified k-fold sampling on the dataset, which divided it into 5 folds to maintain a balanced representation of organs across the training and validation sets. Following this, we resized all HPA images by scaling them down from their original resolution of 3000x3000 to a smaller size of 768x768. This resizing step aimed to reduce computational load while preserving essential image details for accurate segmentation.  The Vahadane stain normalization method \cite{vahadane} was used to minimize discrepancies in image appearance due to staining variations present across HPA and HuBMAP data. To improve the dataset's quality and increase the dataset, we used various image augmentation techniques. This involved applying spatial transformations such as random cropping, flipping (horizontally and vertically), and rotations, which allowed the model to adapt to different geometric variations. In addition, we added artificial noise to simulate any potential acquisition artifacts or noise that may be present in real-world data. To augment the model's learning capabilities, we applied color adjustments such as Hue Saturation Value (HSV), brightness, and contrast to introduce variations in color conditions, and finally, elastic transformation and grid distortion techniques were utilized.

During model training, a batch size of 4 was chosen for the training epoch and 8 for the validation epoch to accommodate hardware limitations. The model was trained for 120 epochs with a learning rate of $5\mathrm{e}{-5}$, with the ReduceLROnPlateau scheduler.
 Incorporating an auxiliary loss with BCE Loss into the training process aimed to promote better learning and regularization. In the initial stage of the model, the auxiliary loss was added from the second encoder block. After 60 epochs of training, the BCE Loss + auxiliary loss was switched to target the first encoder block and trained for another 60 epochs. This strategy encouraged the model to capture informative features, ultimately enhancing segmentation.

\section{6. Inference Pipeline}\label{sec_6}

\begin{figure}[h]%
\centering
\includegraphics[width=0.9\textwidth]{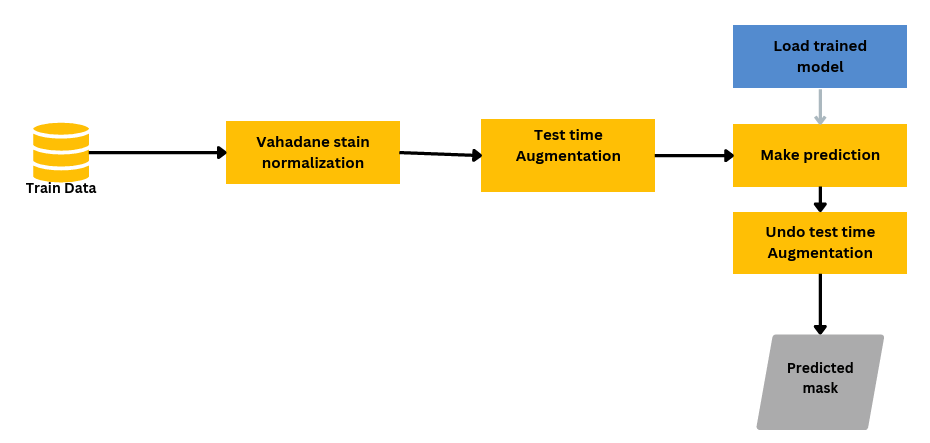}
\caption{Inference Framework}
\label{fig:fig_infer}
\end{figure}

\begin{figure}[h]%
    \centering
    \subfloat[\centering  Image]{{\includegraphics[width=4cm]{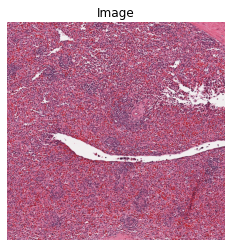} }}%
    \qquad
    \subfloat[\centering Stain normalized]{{\includegraphics[width=4cm]{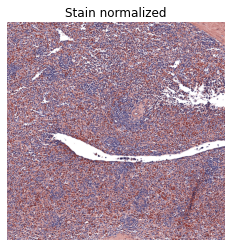} }}%
    \caption{Sample HuBMAP test image with and without stain normalization}%
    \label{fig:infer_stain}%
\end{figure}
The results of the analysis were obtained from a model evaluation that was performed on a single fold. To ensure the good accuracy and consistency of the results, specific threshold values were used for generating masks across both data. For the HPA images, a threshold of 0.5 was set, while for the HuBMAP images, a threshold of 0.4 was used, except for lung samples. Due to the comparatively lower segmentation quality for lung samples, a reduced threshold of 0.15 for HPA and 0.1 for HuBMAP was considered appropriate.

During inference process, the test images were resized to a standardized dimension of 768 x 768 pixels. After that, a stain target image (as shown in Figure \ref{fig:infer_stain}) was used to perform stain normalization, which ensured consistency in image appearance across different samples. Additionally, some test-time augmentation techniques such as horizontal and vertical flips were also used.The trained model then receives this processed data, and predictions are produced according to the organ-specific threshold values. This approach facilitated the generation of final masks, precisely outlining the regions of interest within the medical images.

\section{Results}\label{sec5}

\begin{figure}[!h]
\centering
\begin{tabular}{cccc}
\subfloat{\includegraphics[width = 1.5in]{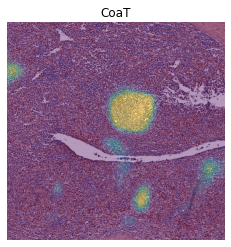}} &
\subfloat{\includegraphics[width = 1.5in]{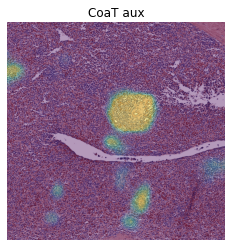}} &
\subfloat{\includegraphics[width = 1.5in]{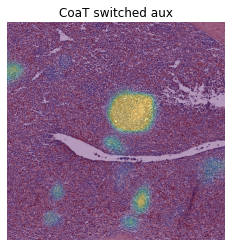}} \\
\subfloat{\includegraphics[width = 1.5in]{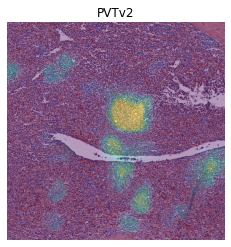}} &
\subfloat{\includegraphics[width = 1.5in]{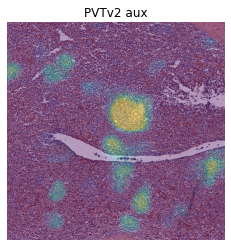}} &
\subfloat{\includegraphics[width = 1.5in]{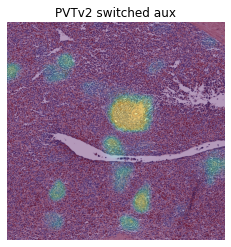}} \\
\subfloat{\includegraphics[width = 1.5in]{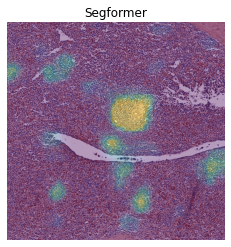}} &
\subfloat{\includegraphics[width = 1.5in]{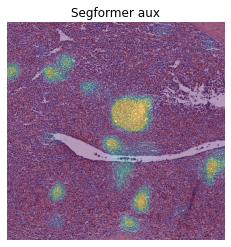}} &
\subfloat{\includegraphics[width = 1.5in]{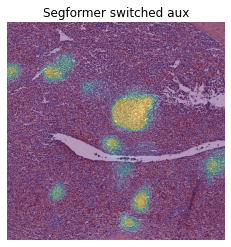}}
\end{tabular}
\caption{Results of different models on sample HuBMAP test image of spleen}
\label{fig:results}
\end{figure}

This section presents the quantitative and qualitative evaluations of the models trained with different loss functions. Table \ref{tab:dice score} summarizes the Dice scores obtained by various models trained on different datasets and loss functions. These models include transformer-based architectures equipped with a DaFormer decoder, such as CoaT, PVTv2, and Segformer, as well as CNN-based architectures like Enet-b5 Unet, Resnet101 Unet, Enet-b5 DLv3+, and Resnet101 DLv3+.

The results demonstrate the impact of incorporating auxiliary loss functions on model performance. Across both public and private datasets, the models trained with a single auxiliary loss exhibit a slight decrease in the Dice score compared to those trained without auxiliary loss. For instance, when using auxiliary loss in the training of the CoaT model on the public dataset, a decrease of about 2.5\% in the Dice score is observed compared to the normal training approach. Similar trends are also observed with other models. However, a switched auxiliary loss technique shows interesting results. In this technique, the auxiliary loss is moved from one block to another, which changes the training dynamics.
Interestingly, all transformer-based models show an improvement in Dice score when trained with switched auxiliary loss, achieving approximately 1\% higher score over both the baseline and single auxiliary loss. This improvement is consistent across both public and private datasets, indicating the effectiveness of this training strategy in enhancing model generalizability and performance. The performance of CNN models on public dataset is evaluated and they undergo training with different loss functions, similar to transformers. When comparing the performance of CNN models to transformer-based architectures, both are sensitive to the use of auxiliary loss. However, the magnitude of the performance impact differs. CNN models tend to display slightly less sensitivity to the addition of auxiliary loss, with smaller variation in Dice scores observed across different training methodologies.

\begin{figure}[!h]
\centering
\begin{tabular}{cccc}
\subfloat{\includegraphics[width = 2.3in]{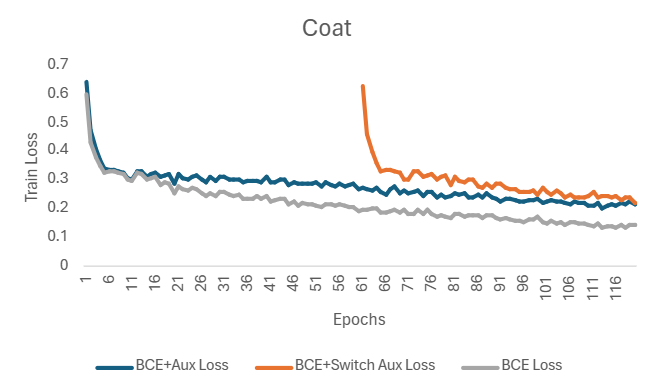}} &&
\subfloat{\includegraphics[width = 2.3in]{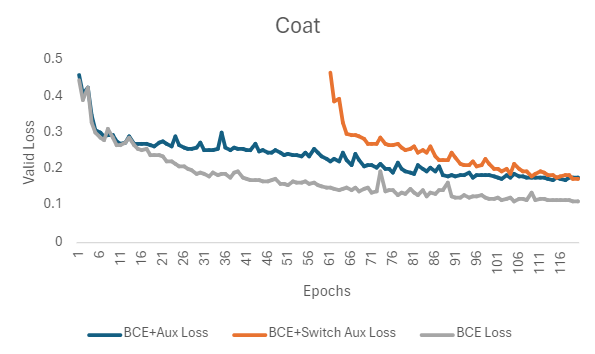}} \\
\subfloat{\includegraphics[width = 2.3in]{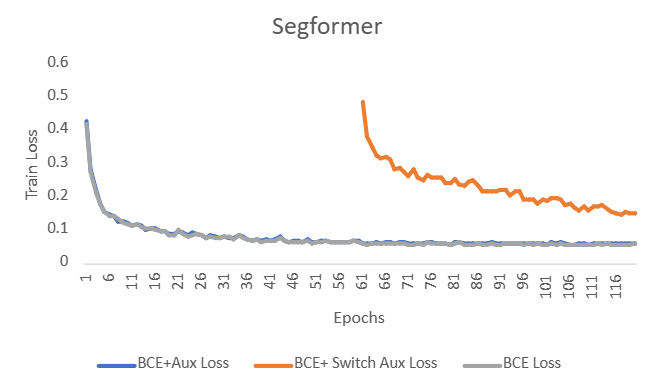}} &&
\subfloat{\includegraphics[width = 2.3in]{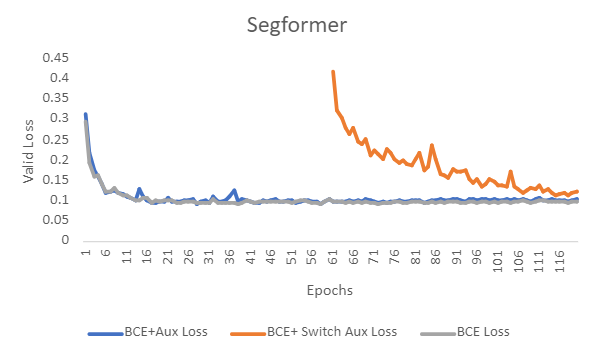}} \\
\subfloat{\includegraphics[width = 2.3in]{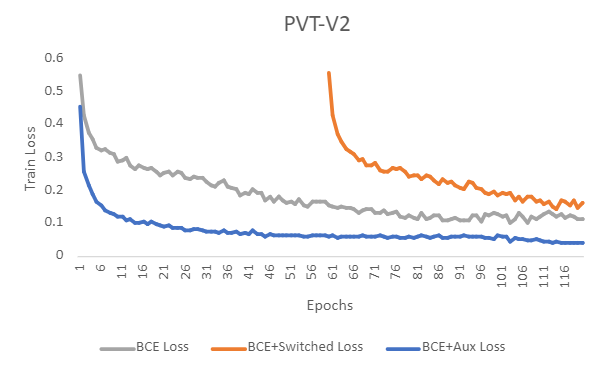}} &&
\subfloat{\includegraphics[width = 2.3in]{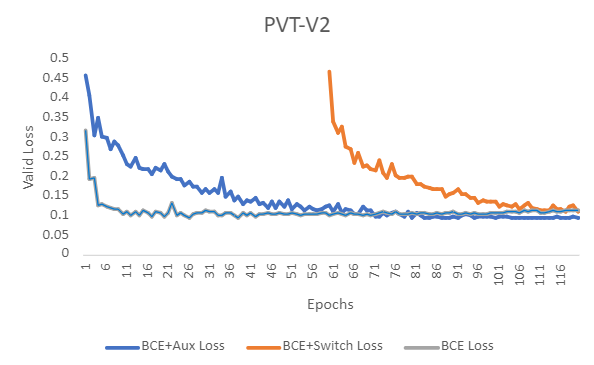}}
\end{tabular}
\caption{Training and Validation Loss Comparison for Three Models}
\label{fig:train_graph}
\end{figure}

\begin{figure}[!h]
\centering
\begin{tabular}{cccc}
\subfloat{\includegraphics[width = 1.3in]{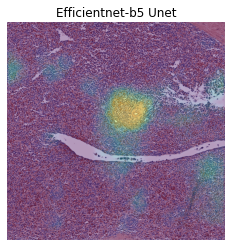}} &
\subfloat{\includegraphics[width = 1.3in]{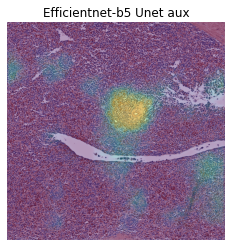}} &
\subfloat{\includegraphics[width = 1.3in]{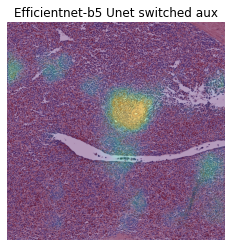}} \\
\subfloat{\includegraphics[width = 1.3in]{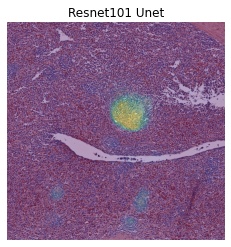}} &
\subfloat{\includegraphics[width = 1.3in]{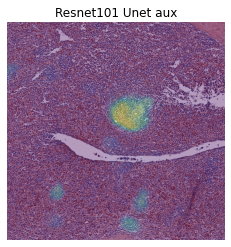}} &
\subfloat{\includegraphics[width = 1.3in]{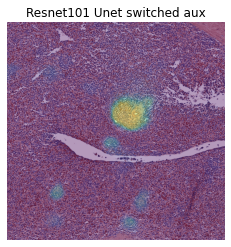}} \\
\subfloat{\includegraphics[width = 1.3in]{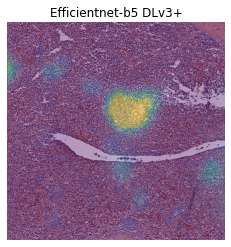}} &
\subfloat{\includegraphics[width = 1.3in]{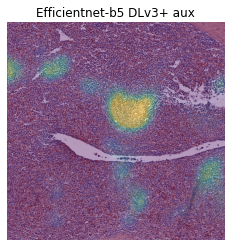}} &
\subfloat{\includegraphics[width = 1.3in]{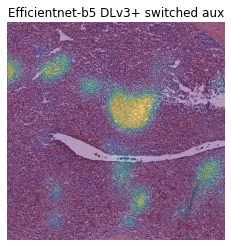}} \\
\subfloat{\includegraphics[width = 1.3in]{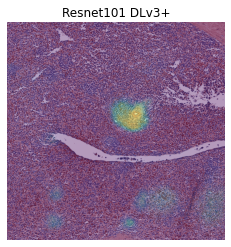}} &
\subfloat{\includegraphics[width = 1.3in]{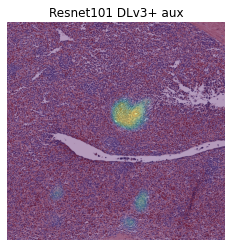}} &
\subfloat{\includegraphics[width = 1.3in]{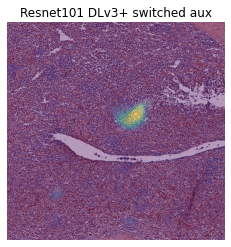}}
\end{tabular}
\caption{Results of CNN models on sample HuBMAP test image of spleen}
\label{fig:cnn_results}
\end{figure}

Figure \ref{fig:results}, \ref{fig:cnn_results} illustrates the predicted mask overlays on a sample HuBMAP test image generated by both transformer-based models and CNN models, trained on various loss functions. We have observed that the Coat model, with switched auxiliary loss, is effective in identifying specific areas of interest, especially the white pulp FTU region. On the other hand, some inconsistencies were noticed in the segmentation outputs of the remaining models, in certain regions. Nevertheless, despite these minor discrepancies, both transformer-based and CNN models show promising performance in capturing the anatomical structures of interest within the HuBMAP test image. The graph depicted in Figure \ref{fig:train_graph} illustrates the training and validation loss curves for all three transformer models.
\begin{table}[htbp]
\centering
\caption{Public and Private HUBMAP and HPA Dataset (dice Score)}
\label{tab:dice score}
\begin{tabular}{@{}p{3cm}p{3cm}p{1.5cm}p{1.5cm}p{1.5cm}@{}}
\toprule
\textbf{Model} & \textbf{Dataset} & \textbf{Normal} & \textbf{With \newline Aux Loss} & \textbf{Switched Aux Loss} \\ \midrule
Enet-b5 Unet & [Pub] HuBMAP+ HPA (CNN) & \textbf{0.77197} & 0.76929 & 0.75352 \\
Resnet101 Unet & [Pub] HuBMAP+ HPA (CNN) & 0.69737 & \textbf{0.70302} & 0.67994 \\
Enet-b5 DLv3+ & [Pub] HuBMAP+ HPA (CNN) & 0.75404 & 0.75438 & \textbf{0.75921} \\
Resnet101 DLv3+ & [Pub] HuBMAP+ HPA (CNN) & 0.60265 & 0.65143 & \textbf{0.66397} \\ 
CoaT & [Pub] HuBMAP+ HPA & 0.78446 & 0.77024 & \textbf{0.79321} \\
PVTv2 & [Pub] HuBMAP+ HPA & 0.74261 & 0.75728 & \textbf{0.75742} \\
Segformer & [Pub] HuBMAP+ HPA & 0.76105 & 0.75931 & \textbf{0.76628} \\
CoaT & [Priv] HuBMAP & 0.76127 & 0.74223 & \textbf{0.77805} \\
PVTv2 & [Priv] HuBMAP & 0.69732 & 0.70284 & \textbf{0.71199} \\
Segformer & [Priv] HuBMAP & 0.70640 & 0.69562 & \textbf{0.72102} \\
\bottomrule
\end{tabular}
\end{table}



\section{Conclusion}\label{sec6}
When dealing with powerful encoders like transformers, it is crucial to achieve optimal training. During backpropagation, it is critical to ensure a smooth gradient flow from the last layers. In our study, we introduced the switched auxiliary loss to improve model convergence. By intentionally making the problem more challenging for the model, we aimed to promote larger loss gradients during backpropagation. This approach facilitated improved model convergence. 
In our experiment using the HuBMAP + HPA: Hacking the Human Body competition dataset, we showed that our proposed approach effectively improves the performance of lightweight transformer models when used for dense prediction tasks in the medical field. Specifically, our results demonstrated that the switched auxiliary loss approach consistently improved performance compared to baseline training and single auxiliary loss approaches. All three transformer models achieved a ~1\% improvement in Dice score on both public and private datasets when trained with the shifted auxiliary loss. The CoaT model performed the best in the task with a score of 0.778 followed by Segformer and PVTv2. According to our findings, using the switched auxiliary loss strategy can improve the generalizability and robustness of transformer-based models for medical image segmentation tasks. This insight is not limited to this specific study and can be applied by practitioners to enhance model training and strengthen robustness in various domains. Ability to train transformer models for images at higher resolutions also remains a challenge due to compute. As we continue to explore the capabilities of transformers, refining training strategies remains a crucial avenue to advance state-of-the-art performance.
\section{Appendix}\label{sec6}
In addition to exploring the impact of auxiliary loss on transformer-based models, we also conducted experiments to assess its effects on models based on convolutional neural networks. The results of the experiments conducted on CNN-based models are presented in Figure \ref{fig:cnn_results}. The findings summarized in Table \ref{tab:dice score} show that incorporating auxiliary loss in the training process leads to improved model performance compared to the standard training approach. However, we observed that the introduction of switched auxiliary loss did not result in significant improvements. This observation highlights the need for further investigation into the effectiveness of switched auxiliary loss in the context of CNN-based architectures.


\bibliography{sn-bibliography}


\end{document}